\documentclass[referee]{raa}           

\usepackage{graphicx,times}
\usepackage{natbib}
\usepackage{amssymb,amsmath}
\usepackage{CJK}
\usepackage[a4paper,margin=1in]{geometry}

\bibpunct{(}{)}{;}{a}{}{,}

\usepackage[pagebackref=true]{hyperref}

\begin{document}

\begin{CJK*}{UTF8}{gkai}

   \title{Search for binarity in Asymptotic Giant Branch stars
   utilizing the future Chinese Space Station Telescope (CSST) }

 \volnopage{ {\bf 20XX} Vol.\ {\bf X} No. {\bf XX}, 000--000}
   \setcounter{page}{1}

   \author{Zhi-Meng Li (李至萌) 
   \inst{1}, Yong Zhang (张泳)\inst{1,2}
   }

  \institute{School of Physics and Astronomy, Sun Yat-sen University, Tang Jia Wan, Zhuhai, 519082, P. R. China; {\it lizhm53@mail2.sysu.edu.cn; zhangyong5@mail.sysu.edu.cn}\\
           \and
                CSST Science Center for the Guangdong-Hong Kong-Macau Greater Bay Area\\
}

\vs \no
   {\small Received 20XX Month Day; accepted 20XX Month Day}

\abstract{Binary systems in the Asymptotic Giant Branch (AGB) phase are widely recognized as a leading theoretical framework underpinning the observed asymmetric morphologies of planetary nebulae. However, the detection of binary companions in AGB systems is severely hampered by the overwhelming brightness and  variability of the evolved primary star, which dominate the photometric and spectroscopic signatures.
Ultraviolet (UV) excess emission has been proposed as a candidate diagnostic for the presence of binary companions in AGB systems.
This paper evaluates the Chinese Space Station Telescope's (CSST) ability to detect UV excess emission in AGB stars, 
leveraging its unprecedented UV sensitivity and wide-field survey capabilities.
We employed synthetic spectral libraries of M0--M8 type giants for primary stars and the ATLAS~9 atmospheric model grid for companion stars spanning a temperature range of 6,500~K to 12,000~K. By convolving these model spectra with the CSST multi-band filter system, we computed color-color diagrams (g-y versus NUV-u) to construct a diagnostic grid.
This grid incorporates interstellar extinction corrections and establishes a framework for identifying AGB binary candidates through direct comparison between observed photometry and theoretical predictions. 
Furthermore, we discuss the physical origins of UV excess in AGB stars. This study pioneers a diagnostic framework leveraging CSST's unique multi-band UV-visible synergy to construct color-color grids for binary candidate identification, overcoming limitations of non-simultaneous multi-instrument observations.
}

   \authorrunning{Li \& Zhang }            
   \titlerunning{Binarity in AGB stars using the CSST}  
   \maketitle

\section{Introduction}           
\label{sect:intro}

The morphological evolution from spherically symmetric asymptotic giant branch (AGB) stars to geometrically complex planetary nebulae (PNe), particularly those exhibiting bipolar and multipolar morphologies, remains a fundamental challenge in late-stage stellar evolution theory.
 Early models proposed toroidal magnetic fields in AGB stars as a fundamental shaping mechanism \citep{Blackman+etal+2001}, wherein poloidal field components compress equatorial mass loss into bipolar outflows. This framework subsequently gained observational support from detections of kilo-Gauss magnetic fields in the central regions of PNe \citep{Jordan+etal+2005}. 
However, \cite{Soker+etal+2006} showed that the angular momentum and energy required to sustain such strong global magnetic fields exceed the limits of the single-star evolution. The bipolar and multipolar geometries of PNe cannot be  attributed solely to the collimating action of magnetic fields within the framework of single-star model.
As a matter of fact, during the thermal pulsation phase of single AGB stars, significant angular momentum is lost due to intense mass loss, which impedes the maintenance of magnetic fields of sufficient strength to shape PNe \citep{Garc_a_Segura_+etal+2014}.

The presence of a binary companion provides a compelling solution to this issue. Gravitational interactions between the AGB star and its companion can replenish angular momentum and trigger asymmetric mass loss through mechanisms such as common envelope evolution (CEE) \citep{Ivanova+etal+2013, Garc_a_Segura_+etal+2021}, grazing envelope evolution \citep{Soker+etal+2015}, and accretion-driven effects \citep{Blackman+etal+2014}. These processes disrupt spherical symmetry, shaping the outflow into distinct nebular structures. Critical parameters, including the remnant core’s effective temperature, the duration of CEE, and the binary system’s mass ratio, further govern the resulting morphology by modulating ejection dynamics and ionization feedback \citep{Garc_a_Segura_+etal+2021}.

Therefore, the identification and characterization of binary AGB stars constitute a critical frontier in resolving the origin of
multipolar morphologies of PNe. 
Conventional binary detection methods, including direct imaging, photometric variability monitoring, and radial velocity measurements,
face severe limitations when applied to pulsating AGB stars due to their high luminosities, and stochastic photospheric oscillations.
 This has driven the development of other detection techniques, such as
 observations of asymmetric CSEs and proper motion variations 
 \citep{Mayer+etal+2012, Pourbaix+etal+2003, Wielen+etal+1996}.

The ultraviolet (UV) excess has long served as a key observational tracer for identifying binary systems, as its presence signals unique evolutionary processes that induce spectral deviations from single-star predictions. These deviations are often linked to dynamical interactions in binary systems. A prominent example is the model proposed by \cite{Han+etal+2007}, which attributes UV excess in elliptical galaxies to enhanced production of hot subdwarf stars through binary mass transfer and merger pathways.
\cite{Sahai+etal+2008} utilized the Galaxy Evolution Explorer (GALEX) to conduct a survey of 21 AGB stars in the far-ultraviolet (FUV) and near-ultraviolet (NUV) bands, with focused analysis on four oxygen-rich sources exhibiting high signal-to-noise ratios in both spectral bands.
Given that the photospheric temperature cutoff of AGB  stars produces negligible UV emission shortward of 2800\,{\AA},
 these excesses likely trace main-sequence companions. 
 This establishes UV excess as a robust binary diagnostic, circumventing limitations of single-epoch optical observations.
The prototypical binary system V Hya exemplifies the power of this methodology. Multi-epoch optical spectroscopy \citep{Planquart+etal+2024} confirms orbital motion, while interferometric observations \citep{Sahai+etal+2009,Sahai+etal+2022a} reveal collimated bipolar outflows shaped by binary interactions. The UV photons from the hot companion drive unique photochemistry in the CSE, altering molecular abundances \citep[see e.g.][]{Sahai+etal+2018}.

The historical detection of UV excess in AGB stars has been fundamentally constrained by the observational limitations of the GALEX, which exclusively operated in the UV bands. 
Analysis of these systems requires synergistic multi-wavelength data spanning UV to near-infrared regimes to disentangle companion-induced UV excess from stochastic chromospheric activity, calibrate dust extinction effects, and obtain a precise spectral energy distributions (SEDs).
The studies replying on the GALEX data inherently suffer from cross-catalog systematic errors due to non-simultaneous observations, 
photometric zeropoint inconsistencies, and aperture mismatch errors.
The Chinese Survey Space Telescope (CSST), a significant component of China's Manned Space Engineering, covers 255--1,000,nm (NUV to near-infrared), enabling simultaneous multi-band observations to minimize cross-instrument errors \citep{Ye+etal+2018}.

In this paper,
we develop a systematic diagnostic framework for identifying AGB binary candidates through UV-excess signatures.
serve as a valuable tool for upcoming CSST surveys.
Section \ref{sect:data} presents the synthetic stellar spectra and the CSST's photometric configurations.
Section \ref{sect:methodology} details the methodology employed for deriving the color-color diagram and constructing the diagnostic grid.
In Section \ref{sect:analyze}, we assess the ability of CSST to 
detect  binary AGB candidates in terms of the diagnostic grid.
 In Section \ref{sect:discussion}, we discuss other potential origins
 of the UV excess. Section \ref{sect:conclusion} presents the conclusions.

\section{Theoretical spectra and CSST bands}
\label{sect:data}

\begin{figure}
   \centering
  \includegraphics[width=0.9\textwidth, angle=0]{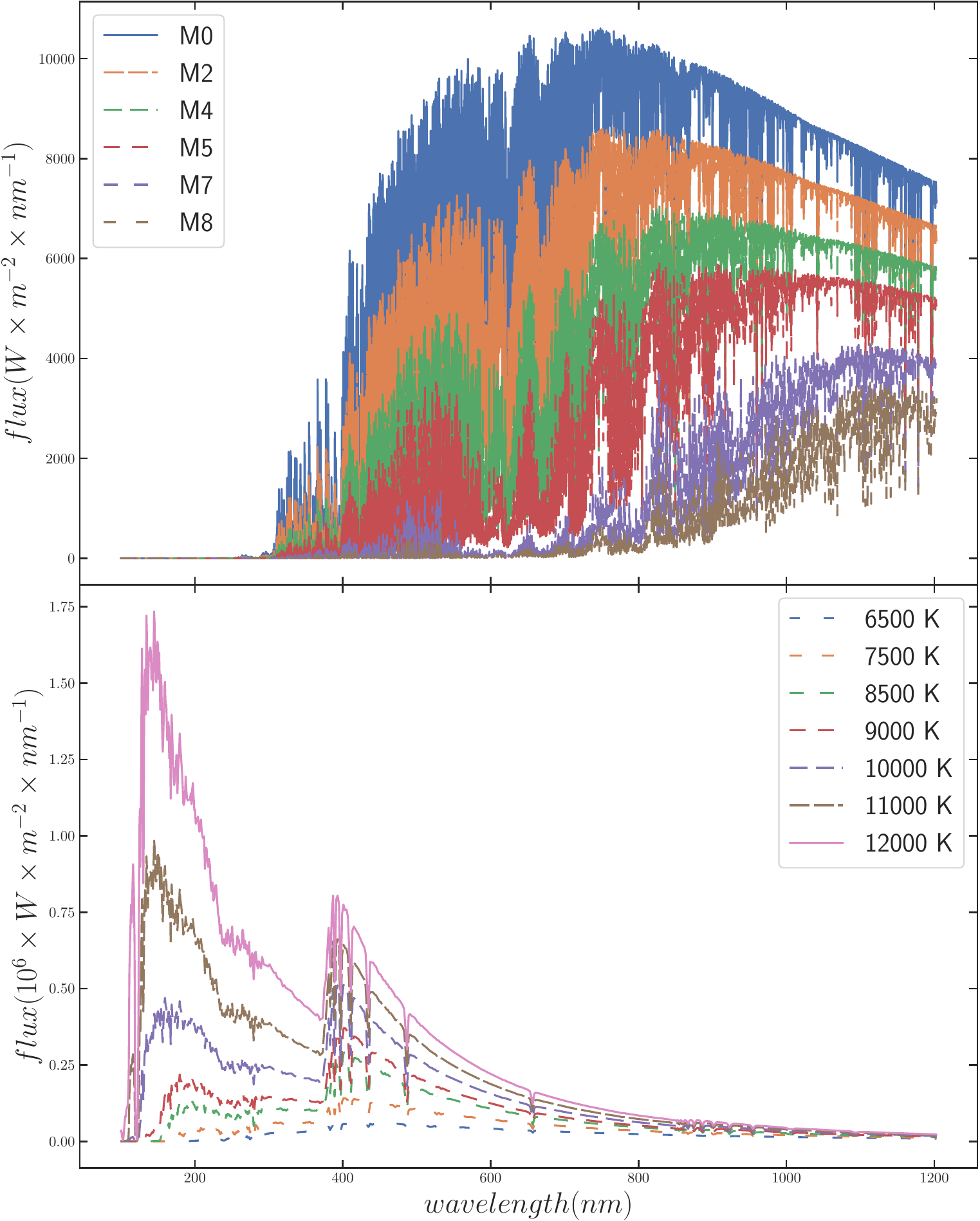}
   \caption{Upper panel: Synthetic spectra for M0--M8 giant stars. Lower panel: Theoretical spectra of main-sequence stars at $6500$\,K-$12000$\,K, calculated from the ATLAS~9 stellar atmosphere models.} 
   \label{Fig1}
\end{figure}

The simulated spectra for both stellar components are presented in Figure~\ref{Fig1}. For the primary stars, we utilize the M-type giant synthetic spectral library of \citet{Fluks+etal+1994}, which combines Uppsala atmospheric models with observational molecular line data to generate calibrated synthetic spectra across M0--M10 subtypes. Our analysis focuses on the M0--M8 regime where decreasing effective temperatures (3,800--2,900~K) drive progressive molecular line blanketing, substantially modifying continuum morphology through TiO/VO band absorption. 

\cite{Ortiz+etal+2016} employed the FUV-NUV color index derived from GALEX observations to constrain the temperatures of companion stars in a sample of AGB binary systems. These systems were previously identified in prior studies \citep[e.g.,][]{Sahai+etal+2008,Famaey+etal+2009} as hosting main-sequence stellar companions. Specifically, these companions were classified as main-sequence stars of spectral type earlier than K0, with most being A- or B-type stars and exhibiting temperatures ranging from 6,500 K to 12,000 K. Therefore, for the companion stars, we adopt the ATLAS~9 model grid \citep{castelli+etal+2004}, selecting main-sequence models with \( T_{\rm eff} = 6,500\mbox{--}12,000~\mathrm{K} \) (B--A spectral types), solar metallicity, and \(\log g = 4.0~\mathrm{dex}\) \citep{Eker+etal+2018} to compute synthetic spectra. This parameter space captures the radiative interplay between cool evolved giants and hot compact companions in binary systems.
Hotter stars are expected to evolve faster than their primary counterparts, whereas cooler stars are incapable of producing prominent UV emission.

\begin{figure}
   \centering
  \includegraphics[width=\textwidth, angle=0]{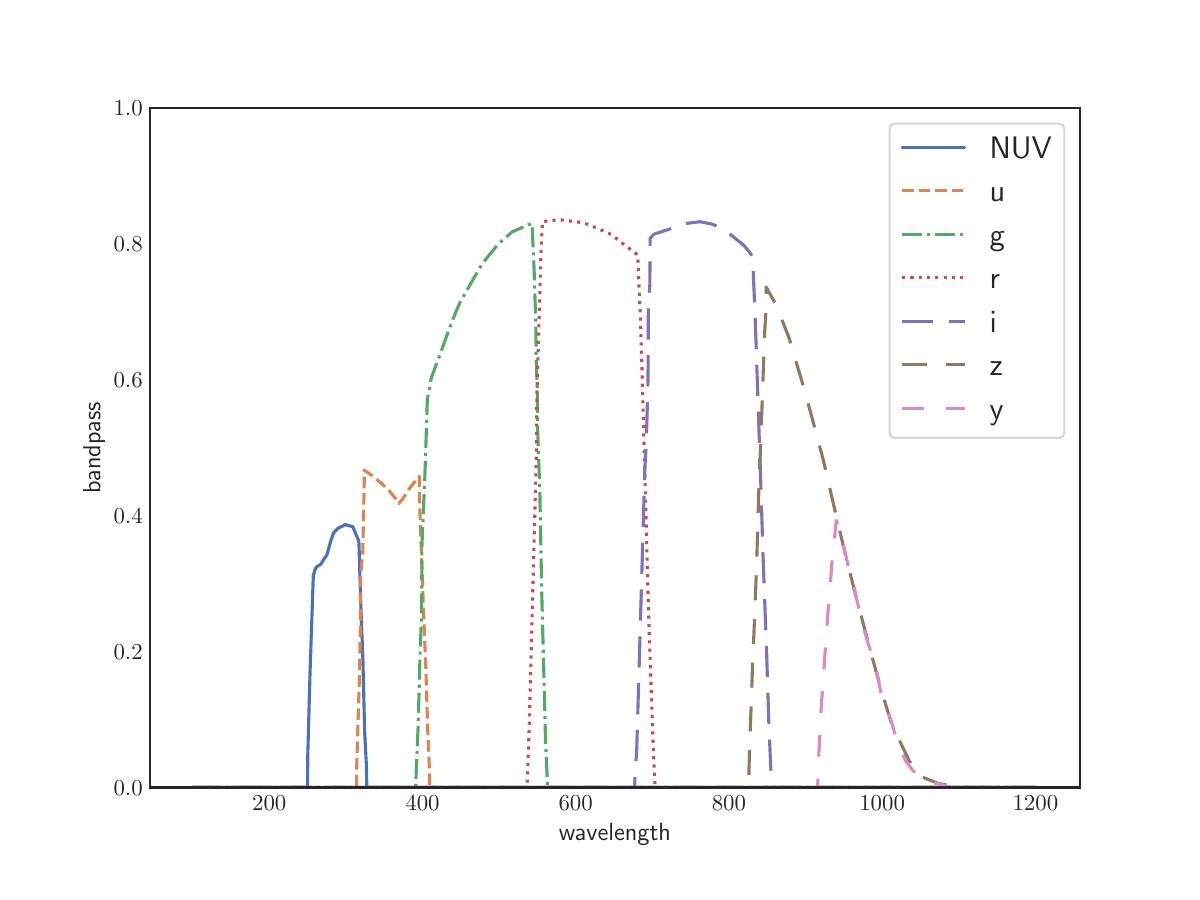}
   \caption{Transmission curves for the 7 filters of CSST.} 
   \label{Fig2}
\end{figure}

The CSST Optical Survey employs a seven-band filter system (NUV, $u$, $g$, $r$, $i$, $z$, $y$) spaning 
a wavelength range from 0.225 to 1.7 $\mu m$.
They facilitate simultaneous three-channel imaging and effectively reduce the cross-calibration uncertainties 
that are typically associated with multi-instrument observations. Table~\ref{tab1} quantifies the filter characteristics: 
mean wavelength ($\lambda_{mean}$), full width at half maximum (FWHM), cutoff wavelengths at 1{\%} ($\lambda_{\pm01}$) and {90}{\%} ($\lambda_{\pm90}$) transmission, peak efficiency ($\eta_{max}$), and {5}{$\sigma$} AB magnitude limits ($m_{lim}$). The corresponding transmission profiles are shown in Figure~\ref{Fig2}.

\section{Methodology}
\label{sect:methodology}

The diagnostic framework makes use of a color - color diagram (g\(-\)y versus NUV\(-\)u) to investigate binary system parameters, taking advantage of the unique sensitivities inherent in these indices. The NUV\(-\)u color index serves as a critical diagnostic tool for investigating UV emission in AGB binary systems. Given that AGB stellar spectra show a sharp cutoff near 2800\,~\AA, while CSST's NUV band spans 2400\(-\)3200\,~\AA, this particular color index provides a sensitive measure of UV excess. The y-band, positioned at the longest wavelength end of CSST's coverage, exclusively traces radiation from the AGB primary. The g\(-\)y color index is adopted as the baseline tracer of the primary's spectral type because it better isolates spectral type variations of the primary component, avoiding potential contamination by molecular absorption(TiO/VO) features that could affect the r\(-\)y, i\(-\)y and z\(-\)y color indices.
Synthetic photometry was produced by convolving atmospheric models of the binary system with CSST transmission curves.
Color index values were derived relative to single-star templates, followed by iterative interstellar extinction correction, thereby constructing a self-consistent diagnostic grid for identifying interacting binaries through multiband photometric signatures.

\subsection{Color indices}

The composite spectral SED of the binary system was derived through flux superposition, incorporating both components' scaled contributions:
\begin{equation}
f_{total}(\lambda)=f_p(\lambda)+\bigg(\frac{R_c}{R_p}\bigg)^2f_c(\lambda),  \label{1}
\end{equation}
where $f_p(\lambda)$ and $f_c(\lambda)$ represnet 
monochromatic fluxes, while $R_p$ and $R_c$ denote
stellar radii of the primary and companion respectively. 
The primary's radius, according to \cite{van_Belle+etal+1999}, follows a well-established empirical relation with spectral type (ST):
\begin{equation}
R_p = \left[4.04 + 9.58 \times 10^{(0.096 \times ST)}\right] R_\odot
\label{eq2}
\end{equation}
applicable across M-type dwarfs (ST = 6--14 corresponding to M0--M8 spectral classifications). 

The companion star was modeled as a main-sequence star with a temperature ranging from 6,500\,K to 12,000\,K. In the selected temperature regime, previous research indicates that there exists a well-defined linear relationship between the temperature and radii of main-sequence stars \citep{Eker+etal+2018}. Leveraging this linear relationship, we calculated the corresponding stellar radii for the pre-selected temperature values.
 
The flux within each band of the CSST was determined by convolving the composite SED with the instrument's transmission curves, as depicted in Fig.~\ref{Fig2}. For a given filter $X$, the flux 
$F_X$ can be computed using the following formula:
\begin{equation}
F_X=\frac{\int_0^\infty f_{total}(\lambda)S_X(\lambda)d\lambda}{\int_0^\infty S_X(\lambda)d\lambda}\label{3}
\end{equation}
where $S_X(\lambda)$ represents the wavelength-dependent transmission efficiency of the filter.
Subsequently, the color indices were computed using the following formula:
\begin{equation}
X - Y=-2.5\log\left(\frac{F_X}{F_Y}\right),\label{4}
\end{equation}
where $F_X$ and $F_Y$ respent the fluxes in bands $X$ and $Y$,
respectively.

\subsection{Extinction arrow}

To account for the affect of interstellar extinction, the galactic extinction curve \cite{Whittet+etal+2022} was incorporated to adjust the model spectra of both the primary and companion stars. The observed flux $F_{\text{obs}}$ can be expressed as a function of the original flux $F_{\text{origin}}$ through the following equation:  
\begin{equation}
F_{\text{obs}} = F_{\text{origin}} \cdot 10^{-0.4A_{\lambda}} \label{5}
\end{equation}
where $A_{\lambda}$ represents the extinction coefficient corresponding to different wavelengths. This coefficient is calculated using the average galactic extinction curve in conjunction with the following equation:
\begin{equation}
\frac{E_{\lambda}-E_B}{E_V - E_B}=R_V\left(\frac{A_{\lambda}}{A_V}-1\right) \label{6}
\end{equation}
In this study, the average interstellar extinction ratio $R_V$ for the galaxy was set to 3.1.

By comparing the color indices before and after accounting for extinction, an extinction vector was derived. Specifically, on the diagnostic diagram, the color-color points determined by the observed fluxes after extinction deviate from their true values and tend to shift along the extinction vector. Notably, in practical scenarios, circumstellar dust can alter the extinction curve when the size of dust grains in the envelope differs from that in the interstellar medium (ISM), thereby modifying both the direction and magnitude of the extinction vector.

\section{Results and analyzes}
\label{sect:analyze}

\subsection{Diagnostic grid}

\begin{figure}
   \centering
  \includegraphics[width=\textwidth, angle=0]{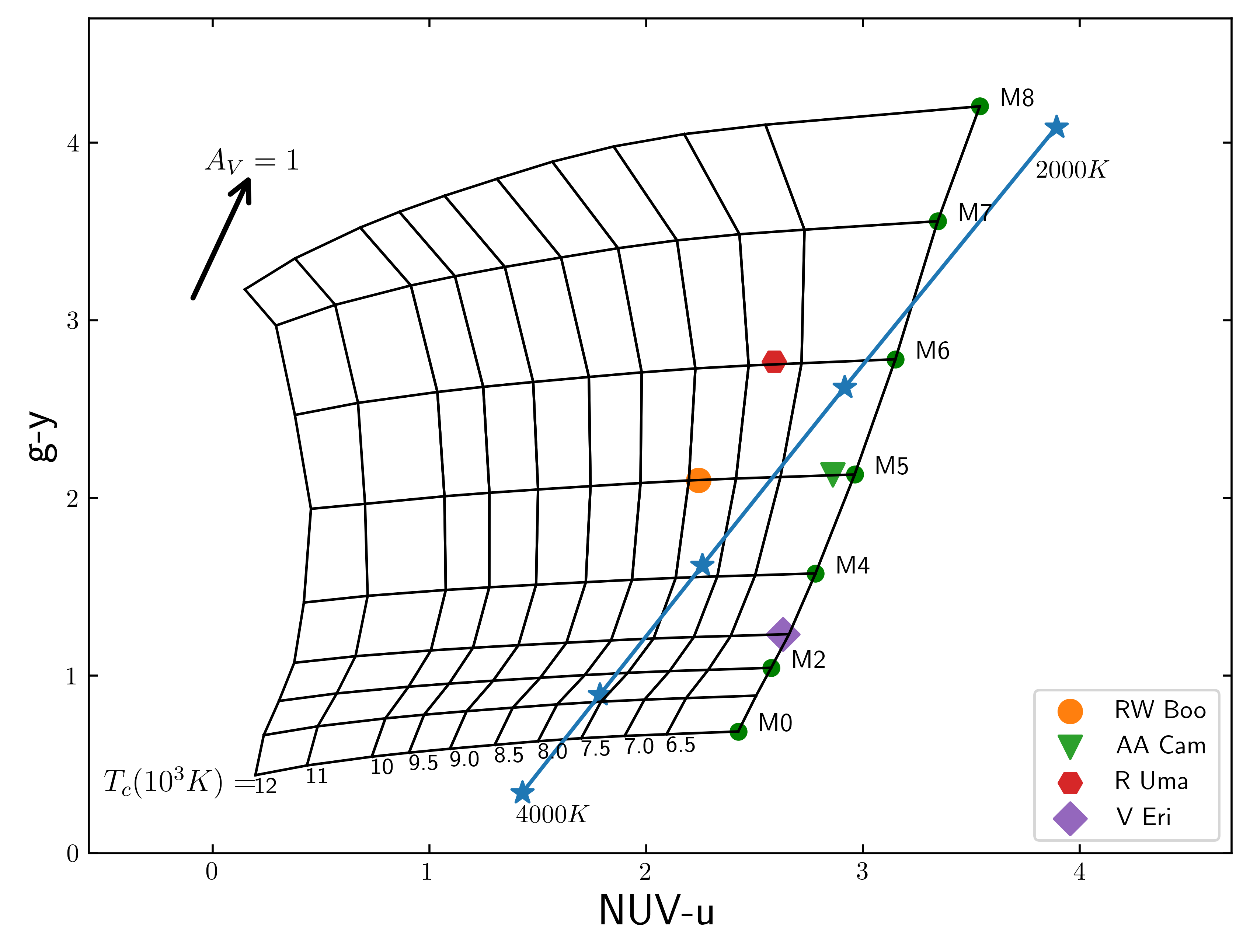}
   \caption{CSST AGB binary diagnostic diagram (g$-$y versus NUV$-$u). The straight line connecting the star symbols indicates blackbody colors corresponding to temperatures ranging from 2,000\,K (upper right) to 4,000\,K (lower left), in 500\,K increments.
    The filled circles represent the colors of M0--M8 type stars. The grid lines represent the colors of binary systems composed of AGB primary stars and main-sequence hot companions with temperatures ranging from 6500\,K to 12000\,K. 
    An extinction vector (black arrow in upper-left corner) denotes the direction and magnitude of interstellar reddening，
   with its length scaled to $A_V=1$.
  } 
   \label{Fig3}
\end{figure}

Figure~\ref{Fig3} showcases the diagnostic grid devised for identifying AGB binary candidates through CSST observations. This grid integrates a color-color diagram 
(g$-$y versus NUV$-$u) and an extinction vector to factor in the effects of interstellar extinction. On the far right of the grid, a series of dots represent the color-color characteristics of AGB primary stars. As the temperature rises, both color index values decline. This is because as temperature increases, short-wavelength radiation experiences a more substantial increase. The g$-$y color indices of the primary stars shows a far stronger dependence on temperature variations compared to the NUV$-$u color indices, which reflects the minimal intrinsic UV emission from AGB stars. For comparison, Figure~\ref{Fig3} also shows color-color characteristics 
of  blackbody emission,  where the temperatures range from 2,000\,K to 4,000,K. 

The  correspondence between spectral types and  $T_{\text{eff}}$ of AGB stars is tabulated in Table~\ref{tab2}. Comparative analysis of these empirical relationships with the blackbody spectral signatures projected onto the diagnostic grid reveals distinct thermal regimes: for M0-type AGB stars ($T_{\text{eff}} = 3,\!895$\,K), the $g$--$y$ color indices yield temperature estimates ($3,\!500\text{--}3,\!900$\,K) that remain approximately consistent with blackbody continuum predictions, whereas 
for an M8 spectral type ($T_{\text{eff}} = 2,\!890$\,K), there is a marked divergence between the temperature derived from the g$-$y color residuals and the blackbody prediction. This  originates from enhanced molecular opacity in cooler AGB atmospheres. As $T_{\text{eff}}$ decreases below $\sim\!3,\!000$\,K, titanium oxide (TiO) bands and other molecular features increasingly dominate the spectral energy distribution, inducing significant deviations from the blackbody radiation.

The grids in Fig.~\ref{Fig3} delineate the colors
of binary system comprising M0--M8 AGB primaries paired with main-sequence companions with $T_{eff}$ spanning $6,\!500\text{--}12,\!000$\,K. The NUV--$u$ color index declines with
increasing companion $T_{eff}$, a trend that becomes increasingly prominent for late-type primaries 
 ($T_{\text{eff}} < 3,\!200$\,K). 
 This behavior arises because UV excess measurements primarily probe the radiative output of hot companions rather than the intrinsic photospheric emission of the AGB stars.

The extinction vector positioned in the upper-left corner of Fig.~\ref{Fig3} demonstrates interstellar dust effects under parameters ($R_V = 3.1$, $A_V = 1$).
While this vector provides insight into dust-induced spectral modifications, actual observational constraints require customized $R_V$ and $A_V$ adjustments depending on the target's Galactic coordinates and sightline dust properties.

The diagnostic framework operates under the paradigm that UV excess in AGB systems predominantly signals main-sequence companions. Through iterative dereddening procedures and grid positioning, this method enables simultaneous determination of primary spectral types and companion temperatures. Systematic errors primarily arise from  discrepancies between synthetic model atmospheres and actual stellar SEDs, particularly regarding molecular band strengths in late-type AGB stars.

\subsection{Observational capability of CSST}

\begin{figure}
   \centering
  \includegraphics[width=0.9\textwidth, angle=0]{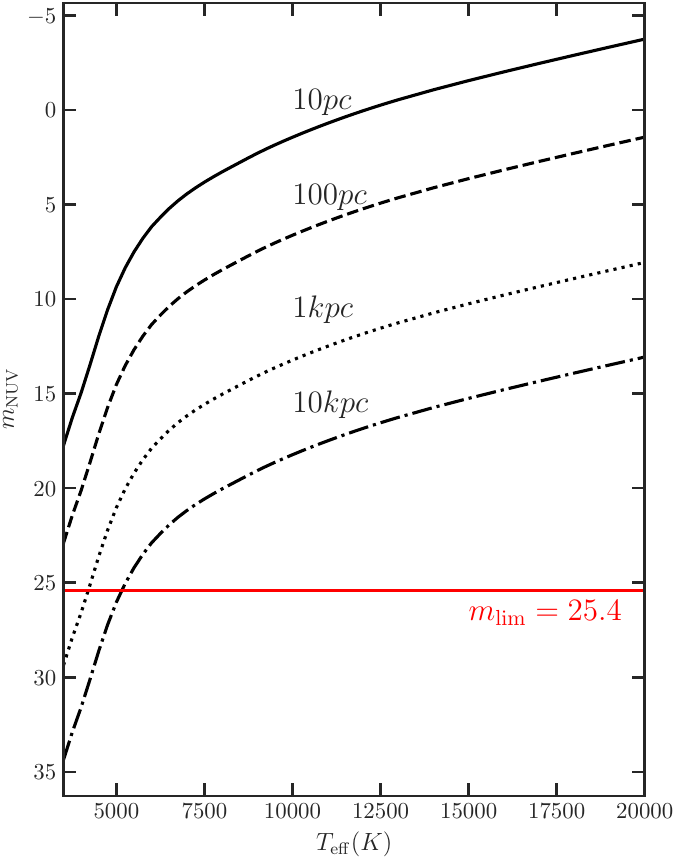}
   \caption{Magnitudes versus effective temperature of   main-sequence stars observed by CSST at different distances. For the distances of 10 pc, 100 pc, 1 kpc, and 10 kpc, the values of $A_V$ are set to 0, 0.1, 1, and 1 respectively. The horizontal line corresponds to the observational limit of the CSST, with a magnitude value in the NUV band, $mag_{NUV}$， equal to 25.4. } 
   \label{Fig4}
\end{figure}

The detectability of main-sequence companions in AGB binaries through CSST's NUV channel requires careful evaluation due to the band's heightened sensitivity to interstellar extinction.
Building on the established observational framework of GALEX from \citet{Ortiz+etal+2016}, we calculated CSST's detection limits using the methodology outlined below:

Synthetic spectra spanning $T_{\rm eff}=3,\!000\text{--}20,\!000$\,K were generated using ATLAS~9 main-sequence atmospheric models. These spectra were then scaled to four representative distances (10\,pc, 100\,pc, 1\,kpc, 10\,kpc) with corresponding visual extinction values $A_V=\{0, 0.1, 1, 1\}$, respectively. The NUV magnitudes were calculated via:
\begin{equation}
{\rm mag}_{\rm NUV} = -2.5\log_{10}\left(\frac{F_{\rm NUV}}{F_{{\rm zero},V}}\right)
\label{eq7}
\end{equation}
where $F_{{\rm zero},V}$ represents the zero-point flux calibration. Comparing these magnitudes against CSST's NUV detection limit (${\rm mag}_{\rm NUV}=25.4$) reveals distinct temperature thresholds: complete detectability ($T_{\rm eff}>3,\!000$\,K) within 100\,pc, declining to $T_{\rm eff}\sim4,\!000$\,K at 1\,kpc and $\sim5,\!000$\,K at 10\,kpc (Fig.~\ref{Fig4}). Notably, the diagnostic grid's companion temperature range ($6,\!500\text{--}12,\!000$\,K) remains fully accessible across all sampled distances. For nearby systems exceeding CCD saturation limits, CSST's slitless spectroscopy mode provides an alternative observational pathway with extended dynamic range.

This analysis demonstrates CSST's capability to probe hot companions in AGB systems out to kiloparsec distances.
The quantitative detection thresholds derived here enable optimized target selection for upcoming CSST surveys of evolved binary populations.

\section{Discussion}
\label{sect:discussion}

\subsection{Diagnostic grid validation}

To assess the reliability of our diagnostic grid, we conducted tests using the SED fitting results of four oxygen-rich AGB stars presented in \cite{Sahai+etal+2008}. These stars were observed in both the FUV and NUV bands of the GALEX and were modeled under the atmospheric constraints available at that time. The companion star temperatures (\(T_c\)), primary component spectral types, secondary component luminosities (\(L_c\)), and the associated luminosity ratios (\(L_p/L_c\)) obtained from \cite{Sahai+etal+2008} are listed in Table \ref{tab3}. Assuming that these four observed objects are AGB binary systems and that the UV excess is attributed to the emission from the main-sequence companions, we selected the spectra of the primary stars according to their spectral types and paired them with the ATLAS~9 
main-sequence star atmosphere models corresponding to the companion star temperatures, which enabled us to simulate the binary emissions of these four observed objects. By using the known temperatures and luminosity ratios of the primary and companion stars, we calculated the radius ratios and subsequently determined their color-color (g$-$y versus NUV$-$u) values, which were then plotted on the diagnostic grid we constructed, as shown in Fig.~\ref{Fig3}. The validation sample for our diagnostic grid is restricted to the four AGB stars from Sahai's study, as these are the sole available objects with complete SED fitting and derived parameters ($T_c, L_p/L_c$) necessary for color index calculations.

We estimate the temperature of the companion stars based on our diagnostic grid, as denoted by \(T_{\text{meas}}\) in Table~\ref{tab3}. A clear discrepancy emerges between \(T_{\text{meas}}\) and the values reported by \cite{Sahai+etal+2008}. The temperature assessments for all four observed objects lie outside the error ranges provided in \cite{Sahai+etal+2008}. While the temperature estimate for RW Boo remains marginally consistent with the cited error range (allowing tentative constraints on the companion's temperature), the values for the remaining three objects  are systematically lower than expected. This inconsistency suggests limitations in the underlying assumptions. Notably, AA Cam and R Uma exhibit extremely faint NUV excess signals, emphasizing the critical need for the CSST to achieve unprecedented sensitivity in detecting subtle UV excess features during AGB binary candidate selection. 

The observed discrepancy between measurements from our diagnostic grid and the 
$T_c$	values reported by Sahai may be attributed to circumstellar dust effects. Specifically, when circumstellar dust grains are smaller than those in the ISM, the resulting steeper extinction curve preferentially absorbs UV photons, potentially explaining this deviation. However, due to the current lack of observational constraints on circumstellar dust properties in these systems, we have not integrated this correction into our analysis.

\cite{Sahai+etal+2008} further noted that for V Eri, AA Cam, and R Uma, no plausible adjustments to the companion's temperature or luminosity could reproduce main-sequence-like spectra. This result aligns with our findings, strongly suggesting that the UV excess in these sources likely originates from mechanisms unrelated to a hot companion's photospheric emission. Consequently, our diagnostic grid -- while effective for systems with clear companion signatures -- may not be universally applicable to all AGB stars exhibiting UV excess. This limitation underscores the necessity to systematically investigate alternative origins of UV excess in AGB stars.

\subsection{Alternative origins of UV excess in AGB stars}

To reliably identify viable AGB binary candidates through UV excess diagnostics, it is critical to disentangle competing mechanisms that could mimic or obscure companion-induced signals. Our study  evaluates three primary origins of UV excess in AGB systems and assess their observational implications for the CSST
and compatibility with our diagnostic grid framework.

\subsubsection{Accretion-driven UV emission}

The UV excess is  likely to stem from the accretion activities within the AGB binary systems. During the accretion process, the interactions among matter particles can potentially result in the generation of high-energy photons.
Previous research on the accretion activities of AGB binaries has mainly centered around X-ray observations, as referenced in \cite{Ramstedt+etal+2012}. When conducting X-ray observations of AGB stars, it has been discovered that the spectral peaks of their X-ray emissions exhibit notable similarities to the characteristic features of accretion processes. The objects under study, T Dra and R Uma, are recognized as binary systems in a variety of star catalogs. Consequently, it is hypothesized that accretion activities could be a probable cause of the observed X-ray emissions.

X-ray spectral analysis of AGB stars reveals that the derived plasma temperature ($T_X$) and luminosity ($L_X$) exhibit systematic correlations with accretion-driven processes, strongly supporting a binary-origin scenario for the X-ray emission rather than stellar atmospheric sources. This conclusion is reinforced by the statistical association between X-ray and
FUV emissions in AGB systems, suggesting a shared physical mechanism for both spectral components through accretion-powered energy dissipation \citep{Sahai+etal+2015}. Multi-wavelength observations of the prototypical system Y Gem demonstrate this connection dynamically, where detected 
high-velocity inflows/outflows and synchronized UV photometric variations correlate temporally with X-ray flux modulations. The observed coupling between UV and X-ray variability timescales, accompanied by characteristic kinematic signatures in spectral line profiles, provides definitive evidence that both emission components originate from accretion shocks in the binary environment \citep{Sahai+etal+2018}.

The CSST's  NUV-visible spectral coverage enables robust temporal correlation analyses free from cross-instrument systematic errors, providing a critical advantage in identifying accretion signatures.
The diagnostic grid presented in this work allows us to conveniently identify 
stellar systems exhibiting NUV-$u$ color indices that deviate from companion-star photospheric models as potential accretion-hosting binaries. Future synoptic CSST observations combining time-resolved UV photometry with 
multi-epoch spectroscopic data will quantitatively test whether the observed NUV-$u$ residual patterns align with characteristic accretion-driven variability metrics -- particularly the stochastic flickering and 
quasi-periodic oscillations diagnostic of accretion disk dynamics. This approach will establish a physical connection between UV excess morphology and accretion physics parameters, potentially revealing how circumstellar material dynamics imprint detectable signatures across electromagnetic bands.

\subsubsection{Emission from the chromosphere}

 \cite{Montez+etal+2017} systematically analyzed UV characteristics of 468 AGB stars through GALEX observations, with 316 targets detected in the UV bands. A distinct subgroup exhibits three diagnostic signatures: \romannumeral1) Chromospheric emission lines (Mg\,{\sc ii}, C\,{\sc ii}, Fe\,{\sc ii}) in UV spectra; \romannumeral2) Correlated variability between NUV and optical fluxes; \romannumeral3) Anti-correlation between NUV luminosity and circumstellar envelope density. These collective features strongly suggest that the observed UV excess -- particularly in the NUV band -- originates from chromospheric processes intrinsic to AGB stars rather than extrinsic mechanisms like binary accretion or companion heating. Supporting this interpretation, \cite{Ortiz+etal+2016} conducted high-resolution UV spectroscopy of 20 FUV/NUV-bright AGB stars, demonstrating that the Mg\,{\sc ii} $\lambda$2800 line intensity scales positively with chromospheric temperature ($T_{\rm chrom}$). This temperature-dependent line strengthening mechanism provides spectral confirmation that chromospheric radiative transfer dominates the NUV excess, with turbulent heating processes in extended atmospheres generating both continuum and line emission components.

Our spectral simulations (Section~\ref{sect:data}) currently omit chromospheric contributions, introducing potential ambiguities in distinguishing binary-induced UV excess from intrinsic stellar atmospheric emission. The methodology for identifying AGB binaries through UV diagnostics has been advanced by \cite{Sahai+etal+2022b}, who constructed a $Cloudy$-based model grid incorporating chromospheric emission across parameterized $T_{\rm chrom}$ and circumstellar density  regimes. Their analysis of simulated FUV-to-NUV flux ratios ($f_{\rm FUV}/f_{\rm NUV}$) revealed a positive correlation with increasing $T_{\rm chrom}$, while observational data show clustered $f_{\rm FUV}/f_{\rm NUV}$ values for single AGB stars. Systems exhibiting $f_{\rm FUV}/f_{\rm NUV} \geq 0.06$ (significantly exceeding the stellar population mean) are thus flagged as binary candidates, as this threshold indicates non-chromospheric contributions.
 
To address current limitations, we propose augmenting our diagnostic framework with: (i) empirical chromospheric templates from archival UV spectra, and (ii) the $Cloudy$ simulated emission models. The CSST NUV bandpass ($\lambda=255$--317\,nm) will enable critical emission-line diagnostics through resolution of chromospheric features like Mg\,{\sc ii} $\lambda$2800, allowing equivalent width measurements to disentangle line-dominated chromospheric emission from companion-star continuum components.

\subsubsection{Effect of scattering in circumstellar dust shells}

\cite{Massey+etal+2005} demonstrated that the red supergiant (RSG) KY Cyg exhibits a significant NUV continuum excess when comparing observed spectra with atmospheric models, accompanied by anomalously high visual extinction relative to non-exhibiting counterparts. 
This reveals dust scattering as the dominant mechanism: shorter wavelengths  experience preferential scattering into the line of sight by circumstellar grains. The dust grains in circumstellar environments can deviate significantly from those in the ISM, leading to distinct extinction curves.

The circumstellar environments of AGB stars bear resemblance to those of RSGs. Both are characterized by dust formation processes and the presence of thick dust shells. The dust in these environments causes the absorption and scattering of stellar light. Consequently, it is essential to take into account the potential for circumstellar dust scattering to result in an UV excess. Nonetheless, significant disparities still exist between the circumstellar environments of AGB stars and RSGs in other respects, such as the scale of mass ejection and the chemical composition of the dust. Therefore, it is necessary to explore the correlation between the UV excess in AGB stars and circumstellar dust parameters to ascertain whether the aforementioned situation of dust-induced UV excess occurs.
 
The multi-band coverage of the CSST offers a distinct advantage. By combining the g$-$y residuals with infrared dust tracers, it becomes possible to isolate the effects of scattering. In future research, radiative transfer models of dust environments will be incorporated to quantify this bias and refine the extinction corrections in the diagnostic grid.
If dust is distributed in an edge-on disk, 
the UV excess resulting from dust scattering should display polarization. In contrast, the excess caused by a companion will not bear any polarization signature. Future polarization observations are expected to directly distinguish between these two scenarios.

\subsubsection{Synthesis for CSST diagnostics}

The  determination of UV excess origins in AGB stars necessitates 
multi-wavelength synergy, particularly incorporating X-ray detections to break degeneracies between accretion and chromospheric scenarios. Future advancements will involve developing mechanism-specific UV excess models -- spanning binary accretion shocks, chromospheric heating, and dust scattering processes -- for quantitative evaluation through CSST observations. Critical to this discrimination is CSST's capacity for simultaneous multi-band monitoring that eliminates cross-calibration uncertainties, enabling detection of subtle variability correlations essential for mechanism identification. The telescope's NUV spectral resolution  permits emission-line diagnostics through features like the Mg\,{\sc ii} $\lambda$2800 doublet, crucial for isolating chromospheric contributions. Furthermore, panchromatic spectral energy distributions spanning 255-900\,nm facilitate dust-scattering model validation via wavelength-dependent flux ratios.
By encoding these observational capabilities into a unified diagnostic framework, CSST will not only identify binary candidates through UV excess signatures but also unravel the dominant physics -- whether powered by magnetospheric accretion, atmospheric turbulence, or circumstellar dust interactions -- thereby advancing binary population statistics and circumstellar environment modeling through mechanism-driven classification.

\section{Conclusion}
\label{sect:conclusion}

AGB binary systems play a pivotal role in understanding the morphological evolution of PNe, particularly their multipolar configurations. The detection of these systems remains a critical challenge, as traditional methods are hindered by the intrinsic high luminosity and thermal pulsations of AGB stars. UV excess has thus emerged as a robust tracer of binarity, given that AGB stars inherently lack UV photon production mechanisms -- observed UV excess is most likely to originate from companion-induced emission or accretion processes.  
 
Our methodology combines M0--M8 giant spectral templates with ATLAS~9 main-sequence companion models ($T_{\rm eff} = 6,500$--$12,000$\,K) to simulate composite binary spectra. By convolving these synthetic spectra with CSST filter transmission profiles, we constructed diagnostic g$-$y vs. NUV-$u$ color-color diagrams. This grid enables determination of binary parameters (primary spectral type, secondary $T_{\rm eff}$) for UV-excess AGB systems after extinction correction.  
 
Extinction effects on CSST's detection efficiency were quantified using main-sequence models spanning $T_{\rm eff} = 3,000$--$20,000$\,K, establishing a 10\,kpc detection horizon for companions within our parameter space. However, when applying this framework to four AGB binary candidates from \cite{Sahai+etal+2008}, the derived temperatures show inconsistencies with literature values, revealing limitations in single-mechanism interpretations.  
 
Current evidence suggests three competitive UV excess origins: (i) binary accretion processes, (ii) AGB chromospheric activity, and (iii) circumstellar dust scattering. Disentangling these mechanisms requires systematic simulations of each scenario under CSST's observational constraints. While our diagnostic grid provides a foundational tool, further refinements in extinction modeling and multi-wavelength diagnostics remain essential priorities for CSST's AGB binary surveys.

\normalem
\begin{acknowledgements}

We're sincerely grateful to the anonymous reviewer for his/her insightful suggestions that refined the paper.
The financial supports of this work are from 
the science research grants from the China Manned Space Project (NO. CMS-CSST-2021-A09, CMS-CSST-2021-A10, etc)，the National Natural Science Foundation of China (NSFC, No.\,12473027 and 12333005), and
the Guangdong Basic and Applied Basic Research Funding (No.\,2024A1515010798).

\end{acknowledgements}
  
\bibliographystyle{raa}
\bibliography{main}                                                                              

\newpage

\clearpage
\newpage
\begin{table}
\bc
\begin{minipage}[]{\textwidth}
\caption[]{Wavelengths and limiting magnitudes of CSST bands. Reference: \cite{Ye+etal+2018}\label{tab1}}\end{minipage}
\setlength{\tabcolsep}{1pt}
\small
 \begin{tabular}{ccccccccccccc}
  \hline\noalign{\smallskip}
Filter& $\lambda_{mean}$& FWHM&$\lambda_{-01}$& $\lambda_{-90}$&$\lambda_{+90}$& $\lambda_{+01}$& $Trans.$&$mag_{lim}$\\
 &(~\AA)&(~\AA)&(~\AA)&(~\AA)&(~\AA)&(~\AA)& & \\
  \hline\noalign{\smallskip}
NUV&2877&701&2480&2550&3170&3260&$65\%$&25.4\\
u&3595&847&3130&3220&3960&4840&$80\%$&25.5\\
g&4978&1562&3910&4030&5450&5610&$90\%$&26.2\\
r&6186&1417&5380&5540&6840&7020&$90\%$&26.0\\
i&7642&1577&6770&6950&8330&8540&$92\%$&25.8\\
z&9046&2477&8250&8460&10650&11000&$92\%$&25.7\\
y&9654&1576&9140&9370&10650&11000&$92\%$&25.5\\

  \noalign{\smallskip}\hline
\end{tabular}
\ec

\tablecomments{0.86\textwidth}{The CSST NUV, u, g, r, i, z, y magnitudes are given in AB system }
\end{table}
\begin{table}
\bc
\begin{minipage}[]{\textwidth}
\caption[]{Effective temperatures of MK spectral subtypes. Reference; \cite{Fluks+etal+1994}\label{tab2}}\end{minipage}
\setlength{\tabcolsep}{7pt}
\small
 \begin{tabular}{c|cccccccccccc}
  \hline\noalign{\smallskip}
  \hline\noalign{\smallskip}
Sp(MK)& $M0$& $M1$&$M2$& $M3$&$M4$& $M5$& $M6$&$M7$&$M8$\\

  \cline{2-10}\noalign{\smallskip}
$T_{eff}(K)$&3895&3810&3736&3666&3574&3434&3309&3126&2890\\

  \noalign{\smallskip}\hline
\end{tabular}
\ec

\end{table}
\begin{table}
\bc
\begin{minipage}[]{\textwidth}
\caption[]{Results from fitting and diagnostic grid. Reference: \cite{Sahai+etal+2008}\label{tab3}}\end{minipage}
\setlength{\tabcolsep}{5pt}
\small
 \begin{tabular}{ccccccccccccc}
  \hline\noalign{\smallskip}
  \hline\noalign{\smallskip}
Target& primary spectral type& $T_c$&$L_c$& $L_p/L_c$&$T_{meas}$\\
 & &(K)&$(L_\odot)$& &(K)\\
  \hline\noalign{\smallskip}
RW Boo&M5&8200(-500,300)&18(-5,7)&280(-80,110)&7400\\
AA Cam&M5&8200(-400,400)&1.1(-0.6,0.7)&3200(-1300,4500)&...\\
V Eri&M6&10000(-700,4100)&6.2(-2.9,2.2)&910(-240,810)&6800\\
R Uma&M3&9200(-300,1100)&0.85(-0.4,0.2)&5300(-900,4700)&...\\
  \noalign{\smallskip}\hline
\end{tabular}
\ec

\end{table}

\end{CJK*}

\end{document}